\documentclass[twocolumn,a4paper,11pt]{article}

\onecolumn

\title{\sf An Algebraic Programming Style for Numerical Software and its Optimization}

\author{{\sf T.B. Dinesh}\\
        {\sf\em\footnotesize Academic Systems Corporation}\\
        {\sf\em\footnotesize 444 Castro Street, Mountain View, CA 94041, USA}\\
        {\tt\footnotesize T\_Dinesh@academic.com}
        \and  
        {\sf Magne Haveraaen}\\
        {\sf\em\footnotesize University of Bergen}\\
        {\sf\em\footnotesize H{\o}yteknologisenteret, N-5020 Bergen, Norway}\\
        {\tt\footnotesize Magne.Haveraaen@ii.uib.no}\\
        \and
        {\sf Jan Heering}\\
        {\sf\em\footnotesize CWI}\\
        {\sf\em\footnotesize P.O. Box 94079, 1090 GB Amsterdam, The Netherlands}\\
        {\tt\footnotesize Jan.Heering@cwi.nl}}

\date{}

\begin{document}
\maketitle
\begin{quote}
\footnotesize
        \noindent {\sf 
        ABSTRACT\\
        The abstract mathematical theory of partial differential
        equations (PDEs) is formulated in terms of manifolds, scalar
        fields, tensors, and the like, but these algebraic structures
        are hardly recognizable in actual PDE solvers.  The general
        aim of the Sophus programming style is to bridge the gap
        between theory and practice in the domain of PDE solvers.  Its
        main ingredients are a library of abstract datatypes
        corresponding to the algebraic structures used in the
        mathematical theory and an algebraic expression style similar
        to the expression style used in the mathematical theory.
        Because of its emphasis on abstract datatypes, Sophus is most
        naturally combined with object-oriented languages or other
        languages supporting abstract datatypes.  The resulting source
        code patterns are beyond the scope of current compiler
        optimizations, but are sufficiently specific for a dedicated
        source-to-source optimizer.  The limited, domain-specific,
        character of Sophus is the key to success here. This kind of
        optimization has been tested on computationally intensive
        Sophus style code with promising results. The general approach
        may be useful for other styles and in other application
        domains as well.}

        \noindent {\sf 
        {\em 1991 Computing Reviews Classification System:} D.1.5, D.2.2, J.2}

        \noindent {\sf 
        {\em Keywords and Phrases:} coordinate-free
        numerics, object-oriented numerics, algebraic programming
        style, domain-specific programming style, optimization of
        numerical code}

        \noindent {\sf 
        {\em Note:} Submitted to {\em Scientific Programming},
        special issue on {\em Coordinate-Free Numerics}.
        This research was supported in part by
	the European Union under ESPRIT Project 21871 (SAGA---Scientific
        Computing and Algebraic Abstractions),
	the Netherlands Organisation for Scientific Research (NWO) under the 
	{\em Generic Tools for Program Analysis and Optimization} project,
	and by a computing resources grant from the 
        Norwegian Supercomputer Committee.\\
        Most of the work reported here was done while
        the first author was at CWI, Amsterdam, The Netherlands, and
        during the second author's sabbatical stay at the
        University of Wales, Swansea, with financial support from the
        Norwegian Science Foundation (NFR).}
\end{quote}

\normalsize
\twocolumn

\section{Introduction} \label{sec:INTRO}

The purpose of the Sophus approach to writing partial differential
equation (PDE) solvers originally proposed in \cite{HaveraaenEtAl92}
is to close the gap between the underlying coordinate-free
mathematical theory and the way actual solvers are written.  The main
ingredients of Sophus are:
\begin{enumerate}

\item A library of abstract datatypes corresponding to manifolds,
      scalar fields, tensors, and the like, figuring in the abstract
      mathematical theory.

\item Expressions involving these datatypes written in a
      side-effect free algebraic style similar to the expressions in
      the underlying mathematical theory.

\end{enumerate}
Because of the emphasis on abstract datatypes, Sophus is most
naturally combined with object-oriented languages or other languages
supporting abstract datatypes.  Hence, we will be discussing
high-performance computing (HPC) optimization issues within an
object-oriented or abstract datatype context, using abstractions that
are suitable for PDEs.

Sophus is not simply object-oriented scientific programming, but a
much more structured approach dictated by the underlying mathematics.
The {\em object-oriented numerics} paradigm proposed in
\cite{BudgeEtAl92,WongEtAl93} is related to Sophus in that
it uses abstractions corresponding to familiar mathematical constructs
such as tensors and vectors, but these do not include continuous
structures such as manifolds and scalar fields. The Sophus approach
is more properly called {\em coordinate-free numerics}
\cite{MuntheKaasHaveraaen96}. 
A fully worked example of conventional vs. coordinate-free programming
of a computational fluid dynamics problem (wire coating for Newtonian
and non-Newtonian flows) is given in \cite{GrantEtAl98}.

Programs in a domain-specific programming style like Sophus may need
additional optimization in view of their increased use of expensive
constructs. On the other hand the restrictions imposed by the style
may lead to new high-level optimization opportunities that can be
exploited by dedicated tools.  Automatic selection of high-level HPC
transformations (especially loop transformations) has been
incorporated in the IBM XL Fortran compiler, yielding a performance
improvement for entire programs of typically less than 2$\times$
\cite[p.~239]{Sarkar97}.  We hope Sophus style programming will allow 
high-level transformations to become more effective than this.

In the context of Sophus and object-oriented programming this article
focuses on the following example.  Object-oriented languages encourage
the use of {\em self-mutating} ({\em self-updating}, {\em mutative\/}) objects rather than
a side-effect free algebraic expression style as advocated by Sophus.
The benefits of the algebraic style are considerable. We obtained a
reduction in source code size using algebraic notation vs. an
object-oriented style of up to 30\% in selected procedures of a
seismic simulation code, with a correspondingly large increase in
programmer productivity and maintainability of the code as measured by
the Cocomo technique \cite{Boehm81}, for instance. On the negative
side, the algebraic style requires lots of temporary data space for
(often very large) intermediate results to be allocated and
subsequently recovered.  Using self-mutating objects, on the other
hand, places some of the burden of variable management on the
programmer and makes the source code much more difficult to write,
read, and maintain.  It may yield much better efficiency, however.
Now, by including certain restrictions as part of the style, a precise
relationship between self-mutating notation and algebraic notation may
be achieved.  Going one step further, we see that the natural way of
building a program from high-level abstractions may be in direct conflict
with the way current compilers optimize program code.  We propose a
source-to-source optimization tool, called CodeBoost, as a solution to
many of these problems.  Some further promising optimization
opportunities we have experimented with but not yet included in
CodeBoost are also mentioned.  The general approach may be useful for
other styles and other application domains as well.

This paper is organized as follows.  After a brief overview of tensor
based abstractions for numerical programming and their realization as
a software library (Section~\ref{sec:TBA}), we discuss the
relationship between algebraic and self-mutating expression notation,
and how the former may be transformed into the latter
(Section~\ref{sec:OPNOTATION}).  We then discuss the 
implementation of the CodeBoost source-to-source optimization
tool (Section~\ref{sec:CODEBOOST}), and give some further examples of how
software construction using class abstractions may conflict with
efficiency issues as well as lead to new opportunities for
optimization (Section~\ref{sec:SOFTWARESTRUCT}).  Finally, we present
conclusions and future work (Section~\ref{sec:FINAL}).

\section{A Tensor Based Library for Solving PDEs} \label{sec:TBA}

Historically, the mathematics of PDEs has been approached in two
different ways. The solution-oriented approach uses concrete
representations of vectors and matrices, discretisation techniques,
and numerical algorithms, while the abstract approach develops the
theory in terms of manifolds, scalar fields, tensors, and the like,
focusing more on the structure of the underlying concepts than on how
to calculate with them (see \cite{Schutz80} for a good introduction).
 
The former approach is the basis for most of the PDE software in
existence today. The latter has very promising potential for the
structuring of complicated PDE software when combined with template
class based programming languages or other languages supporting
abstract datatypes.  As far as notation is concerned, the abstract
mathematics makes heavy use of overloaded infix operators. Hence,
user-definable operators and operator overloading are further
desirable language features in this application domain. C++
\cite{Stroustrup97} comes closest to meeting these desiderata, but,
with modules and user-definable operators, Fortran 90/95
\cite{AdamsEtAl92,AdamsEtAl97} can also be used.  In its current form
Java \cite{GoslingEtAl96} is less suitable.  It has neither templates
nor user-definable operators.  Also, Java's automatic memory management
is not necessarily an advantage in an HPC setting
\cite[Section~4]{SinghalEtAl97}.  Some of these problems may be
alleviated in Generic Java \cite{BrachaEtAl98}.  The examples in this
article use C++.

\subsection{The Sophus Library} \label{sec:SOPHUS}

The Sophus library provides the abstract mathematical concepts from
PDE theory as programming entities. Its components are based on the
notions of manifold, scalar field and tensor field, while the
implementations are based on the conventional numerical algorithms and
discretisations.  Sophus is currently structured around the following
concepts:
\begin{itemize}
	\item Basic $n$-dimensional mesh structures.  These
	are like rank $n$ arrays (i.e., with $n$ independent indices),
	but with operations like $+$, $-$ and $*$ mapped over all
	elements (much like Fortran 90/95 array operators) as well as the
	ability to add, subtract or multiply all elements of the mesh
	by a scalar in a single operation. There are also operations
	for shifting meshes in one or more dimensions. Parallel and
	sequential implementations of mesh structures can be used
	interchangeably, allowing easy porting between architectures
	of any program built on top of the mesh abstraction.

	\item Manifolds. These represent the physical space where the
	problem to be solved takes place. Currently Sophus only
	implements subsets of $R^n$.
 
	\item Scalar fields. These may be treated formally as
	functions from manifolds to reals, or as arrays indexed by the
	points of the manifold with reals as data elements.  Scalar
	fields describe the measurable quantities of the physical
	problem to be solved. As the basic layer of ``continuous
	mathematics'' in the library, they provide the partial
	derivation operations. Also, two scalar fields on the same
	manifold may be pointwise added, subtracted or multiplied.
	The different discretisation methods provide different designs
	for the implementation of scalar fields. A typical
	implementation would use an appropriate mesh as underlying
	discrete data structure, use interpolation techniques to give
	a continuous interface, and map the $+$, $-$, and $*$
	operations directly to the corresponding mesh operations.  In
	a finite difference implementation partial derivatives are
	implemented using shifts and arithmetic operations on the
	mesh.

	\item Tensors. These are generalizations of vectors and
	matrices and have scalar fields as components. Tensors define
	the general differentiation operations based on the partial
	derivatives of the scalar fields, and also provide operations
	such as componentwise addition, subtraction and
	multiplication, as well as tensor composition and application
	(matrix multiplication and matrix-vector multiplication).
	A special class are the metric tensors. These satisfy certain
	mathematical properties, but their greatest importance in this
	context is that they can be used to define properties of
	coordinate systems, whether Cartesian, axiosymmetric or
	curvilinear, allowing partial differential equations to be
	formulated in a coordinate-free way.  The
	implementation of tensors relies heavily on the arithmetic
	operations of the scalar field classes.
\end{itemize}
A partial differential equation in general provides a relationship
between spatial derivatives of tensor fields representing physical
quantities in a system and their time derivatives. Given constraints
in the form of the values of the tensor fields at a specific instance
in time together with boundary conditions, the aim of a PDE solver is
to show how the physical system will evolve over time, or what state
it will converge to if left by itself.
Using Sophus, the solvers are formulated on top of the coordinate-free
layer, forming an abstract, high level program for the solution of the
problem.

\subsection{Sophus Style Examples}

\begin{figure*}
\begin{center}
\rule{15.8cm}{.5mm}
\footnotesize
\begin{verbatim}
/** returns the mesh circularly shifted {i} positions in dimension {d} */
template<class T> Mesh<T> shift(const Mesh<T> & M, int d, int i);

/** returns the elementwise sum of {lhs} and {rhs}   */
template<class T> Mesh<T> operator+(const Mesh<T>& lhs, const Mesh<T>& rhs);

/** returns the elementwise difference of {lhs} and {rhs}   */
template<class T> Mesh<T> operator-(const Mesh<T>& lhs, const Mesh<T>& rhs);

/** returns the elementwise product of the {lhs} and {r} */ 
template<class T> Mesh<T> operator*(const Mesh<T>& lhs, const real& r);

...
\end{verbatim}
\normalsize
\rule{15.8cm}{.5mm}
\end{center}
\caption[Specification of operators on a mesh template class]{Specification 
of algebraic style operators on a mesh template class.} \label{fig:MESH}
\end{figure*}

\begin{figure*}
\begin{center}
\rule{15.8cm}{.5mm}
\footnotesize
\begin{verbatim}
/** some operations on a scalar field implemented using the finite difference method
*/
class TorusScalarField { 
private:
  Mesh<real> msf();   // data values for each grid point of the mesh
  real delta;         // resolution, distance between grid points

  :

public:

  :

/** 4 point derivation algorithm, computes partial derivative in dimension d */
void uderiv (int d) 
{ Mesh<real> ans = (shift(msf,d,1) - shift(msf,d,-1)) * 0.85315148548241;
       ans = ans + (shift(msf,d,2) - shift(msf,d,-2)) * -0.25953977340489;
       ans = ans + (shift(msf,d,3) - shift(msf,d,-3)) * 0.06942058732686;
       ans = ans + (shift(msf,d,4) - shift(msf,d,-4)) * -0.01082798602277;
       msf = ans * (1/delta);
}  

/** adding scalar field {rhs} to this TorusScalarField */
void operator+=(const TorusScalarField& rhs);
{ msf = msf + rhs;
}  

/** subtracting scalar field {rhs} from this TorusScalarField */
void operator-=(const TorusScalarField& rhs);
{ msf = msf - rhs;
}  

/** multiplying scalar {r} to this TorusScalarField */
void operator*=(const real& r);
{ msf = msf * r;
}  

  :
}
\end{verbatim}
\normalsize
\rule{15.8cm}{.5mm}
\end{center}
\caption[A self-mutating implementation of a partial derivation, ..]{A
class \texttt{TorusScalarField} with self-mutating implementations of a
partial derivation algorithm, a scalar field addition, and a scalar
multiplication algorithm. The code itself is using algebraic notation for
the mesh operations.}
\label{fig:DERIVATION}
\end{figure*}

The algebraic style for function declarations can be seen in
Figure~\ref{fig:MESH}, which shows specifications of some operations
for multidimensional meshes, the lowest level in the Sophus
library. The mesh class is parameterized by a class
\texttt{T}, so all operations on meshes likewise are parameterized by
\texttt{T}. Typical parameters would be a float or scalar field class.
The operations declared are defined to behave like pure
functions, i.e., they do not update any internal state or modify any
of their arguments.  Such operations are generally nice to work with
and reason about, as their application will not cause any hidden
interactions with the environment.

Selected parts of the implementation of a continuous scalar field
class are shown in Figure~\ref{fig:DERIVATION}. This scalar field
represents a multi-dimensional torus, and is implemented using a mesh
class as the main data structure. The operations of the class have
been implemented as self-mutating operations
(Section~\ref{sec:OPNOTATION}), but are used in an algebraic way for
clarity. It is easy to see that the partial derivation operation is
implemented by shifting the mesh longer and longer distances, and
gradually scaling down the impact these shifts have on the derivative,
yielding what is known as a four-point, finite difference, partial
derivation algorithm. The addition and multiplication operations are
implemented using the element-wise mapped operations of the mesh.

The meshes used in a scalar field tend to be very large.  A
\texttt{TorusScalarField} may typically contain between 0.2 and 2MB of
data, perhaps even more, and a program may contain many such
variables. The standard translation technique for a C++ compiler is to
generate temporary variables containing intermediate results from
subexpressions, adding a considerable run-time overhead to the algebraic
style of programming. An implementation in terms of self-mutating
operators might yield noticeable efficiency
gains. For the addition, subtraction and multiplication algorithms of
Figure~\ref{fig:DERIVATION} a self-mutating style is easily obtained. The
derivation algorithm will require extensive modification, such as shown in
Figure~\ref{fig:DERIVATIONmut}, with a marked deterioration in readability
and maintainability as a result.

\section{Algebraic Notation and Self-Mutating Implementation} \label{sec:OPNOTATION}

\subsection{Self-Mutating Operations} \label{sec:MUTOPS}

Let {\tt a}, {\tt b} and {\tt c} be meshes with operators as defined
in Figure~\ref{fig:MESH}. The assignment
\begin{verbatim}
c = a * 4.0 + b + a
\end{verbatim}
is basically evaluated as 
\begin{verbatim}
temp1 = a * 4.0;
temp2 = temp1 + b;
c     = temp2 + a.
\end{verbatim}
This involves the creation of the meshes \texttt{temp1}, \texttt{temp2},
\texttt{c}, the first two of which are temporary.
Obviously, since all three meshes have the same size and the
operations in question are sufficiently simple, repeated use of a
single mesh would have been possible in this case.  In fact,
for predefined types like integers and floats an optimizing C or C++ compiler
would translate the expression to a sequence of compound
assignments\footnote{Not to be confused with the C notion of compound
statement, which is a sequence of statements enclosed by a pair of braces.}
\begin{verbatim}
c = a; c *= 4.0; c += b; c += a,
\end{verbatim}
which repeatedly uses variable {\tt c} to store intermediate results.

We would like to be able to do a similar optimization of the mesh
expression above as well as other expressions involving $n$-ary
operators or functions of a suitable nature for user-defined types as
proposed in \cite{Dinesh92}.  In an object-oriented language, it would
be natural to define self-mutating methods (i.e., methods mutating
{\em this}) for the mesh operations in the above expression. These
would be closely similar to the compound assignments for predefined
types in C and C++, which return a pointer to the modified data
structure.  Sophus demands a side-effect free expression style close
to the underlying mathematics, however, and forbids {\em direct} use
of self-mutating operations in expressions.  Note that with a
self-mutating {\tt +=} operator returning the modified value of its first
argument, the expression {\tt (a += b) += a} would yield $2({a}+{b})$
rather than $(2{a})+{b}$.

By allowing the user to define self-mutating operations and providing
a way to use them in a purely functional manner, their direct use can
be avoided.  There are basically two ways to do this, namely, by means
of wrapper functions or by program transformation. These will be
discussed in the following sections.

\subsection{Wrapper Functions} \label{sec:WRAPPER}

\begin{figure*}  
\begin{center}
\rule{15.8cm}{.5mm}
\footnotesize
\begin{verbatim}
  /** implements the basic mesh operations */
  template<class T> class MeshCode1{
        ...
  public:
        /** circularly shifts {this} mesh {i} positions in dimension {d} */
        void ushift(int d, int i){ ... }

        /** adds {rhs} elementwise to {this} mesh   */
        void operator+=(const MeshCode1<T> & rhs){ ... }

        /** subtracts {rhs} elementwise from {this} mesh   */
        void operator-=(const MeshCode1<T> & rhs){ ... }

        /** multiplies {this} mesh elementwise by {r} */ 
        void operator*=(real r){ ... }
        ...
        }
\end{verbatim}
\normalsize
\rule{15.8cm}{.5mm}
\end{center}
\caption[The use of self-mutating membership operations for a mesh class.]{The 
use of self-mutating membership operations for a mesh class \texttt{MeshCode1}.} \label{fig:MESHmut}
\end{figure*}

\begin{figure*}
\begin{center}
\rule{15.8cm}{.5mm}
\footnotesize
\begin{verbatim}
template<class T> MeshCode1<T> shift(const MeshCode1<T> & MD, int d, int i)
   { MeshCode1<T> C = MD; C.ushift(d,i); return C; }
template<class T> MeshCode1<T> operator+(const MeshCode1<T>& lhs, const MeshCode1<T>& rhs);
   { MeshCode1<T> C = lhs; C += rhs; return C; }
template<class T> MeshCode1<T> operator-(const MeshCode1<T>& lhs, const MeshCode1<T>& rhs);
   { MeshCode1<T> C = lhs; C -= rhs; return C; }
template<class T> MeshCode1<T> operator*(const MeshCode1<T>& lhs, const real& r);
   { MeshCode1<T> C = lhs; C *= r; return C; }
...
\end{verbatim}
\normalsize
\rule{15.8cm}{.5mm}
\end{center}
\caption{Wrapper functions implementing the specification of a mesh using 
\texttt{MeshCode1} operations generated by the SCC preprocessor.} \label{fig:MESHwrapper}
\end{figure*}

Self-mutating implementations can be made available to the programmer
in non-self-mutating form by generating appropriate wrapper functions.
We developed a C++ preprocessor SCC doing this. It scans the source text for
declarations of a standard form and automatically creates wrapper
functions for the self-mutating ones.  This allows the use of an
algebraic style in the program, and relieves the programmer of the
burden of having to code the wrappers manually.

A self-mutating operator {\tt op=} is related to its algebraic analog
{\tt op} by the basic rule
\begin{tabbing} 
{\tt x = y op z;} \,$\equiv$\, {\tt x = copy(y); x op= z;} \` (1)
\end{tabbing}
or, if the second argument is the one being updated,\footnote{This
does not apply to built-in compound assignments in C or C++, but
user-defined compound assignments in C++ may behave in this way.}  by
the rule
\begin{tabbing} 
{\tt x = y op z;} \,$\equiv$\, {\tt x = copy(z); y op= x;} \` (2)
\end{tabbing} 
where $\equiv$ denotes equivalence of the left- and right-hand sides,
{\tt x}, {\tt y}, {\tt z} are C++ variables, and {\tt copy} makes a
copy of the entire data structure. Now, the Sophus style does not
allow aliasing or sharing of objects, and the (overloaded) assignment
operator {\tt x = y} is always given the semantics of {\tt x =
copy(y)} as used in (1) and (2).  Hence, in the context of Sophus (1)
can be simplified to
\begin{tabbing} 
{\tt x = y op z;} \,$\equiv$\, {\tt x = y; x op= z;}    \` (3)\\\\
\end{tabbing}  
and similarly for (2). We note the special case
\begin{tabbing} 
{\tt x = x op z;} \,$\equiv$\, {\tt x op= z;}           \` (4)\\\\
\end{tabbing}
and the obvious generalizations
\begin{tabbing} 
{\tt x = x op e;}   \,$\equiv$\,  {\tt x op= e;}            \`(5)\\\\

{\tt x = e1 op e2;} \,$\equiv$\,  {\tt x = e1; x op= e2;}   \`(6)\\\\
\end{tabbing}
where {\tt e}, {\tt e1}, and {\tt e2} are expressions. 
SCC uses rules such as (6) to obtain purely functional behavior
from the self-mutating definitions in a straightforward way.
Figure~\ref{fig:MESHwrapper} shows the wrappers created by SCC for the
self-mutating mesh operations of Figure~\ref{fig:MESHmut}.  The case
of $n$-ary operators and functions is similar ($n \geq 1$).  We note
that, unlike C and C++ compound assignments, Sophus style
self-mutating operators do not return a reference to the variable
being updated and cannot be used in expressions. This simpler behavior
facilitates their definition in Fortran 90/95 and other languages of
interest to Sophus.

The wrapper approach is superficial in that it does not minimize the
number of temporaries introduced for expression evaluation as
illustrated in Section~\ref{sec:MUTOPS}.  We therefore turn to a more
elaborate transformation scheme.

\subsection{Program Transformation} \label{sec:PROGTRANS}

\begin{figure*}
\begin{center}
\rule{15.8cm}{.5mm}
\footnotesize
\begin{verbatim}
/** some operations on a scalar field implemented using the finite difference
    method
*/
public class ScalarField { 
  MeshCode1 msf();   // data values for each grid point of the mesh
  real delta;   // resolution, distance between grid points

  :

/** 4 point derivation algorithm, computes partial derivative in dimension d */
public void uderiv (int d) 
{ MeshCode1 msa = msf; 
  MeshCode1 msb = msf; 
  MeshCode1 scratch(); 

  msa.ushift(d,1);
  msb.ushift(d,-1);
  scratch = msa; scratch.uminus(msb); scratch.umult(0.85315148548241);
  msf = scratch;

  msa.ushift(d,1);
  msb.ushift(d,-1);
  scratch = msa; scratch.uminus(msb); scratch.umult(-0.25953977340489);
  msf.uplus(scratch);

  msa.ushift(d,1);
  msb.ushift(d,-1);
  scratch = msa; scratch.uminus(msb); scratch.umult(0.06942058732686);
  msf.uplus(scratch);

  msa.ushift(d,1);
  msb.ushift(d,-1);
  scratch = msa; scratch.uminus(msb); scratch.umult(-0.01082798602277);
  msf.uplus(scratch);

  msf.umult(1/delta);
}  
  :
}
\end{verbatim}
\normalsize
\rule{15.8cm}{.5mm}
\end{center}
\caption{Optimized partial derivation operator of class
{\tt TorusScalarField} (Figure~\ref{fig:DERIVATION}).} \label{fig:DERIVATIONmut}
\end{figure*}

Transformation of algebraic expressions to self-mutating form with
simultaneous minimization of temporaries requires a parse of the
program, the collection of declarations of self-mutating operators and
functions, and matching them with the types of the operators and
functions actually used after any overloading has been
resolved. Also, declarations of temporaries have to be added with the
proper type.  Such a preprocessor would be in a good position to
perform other source-to-source optimizations as well.  In fact, this
second approach is the one implemented in CodeBoost with promising
results.

Figure~\ref{fig:DERIVATIONmut} shows an optimized version of the
partial derivation operator of class {\tt TorusScalarField}
(Figure~\ref{fig:DERIVATION}) that might be obtained in this way.  In
addition to the transformation to self-mutating form, an obvious
rule for {\tt ushift} was used to incrementalize shifting of the
mesh.

Assuming the first argument is the one being updated, some further
rules for binary operators used in this stage are
\begin{tabbing}
{\tt x op1= e1 op2 e2;} \,$\equiv$\,  \\
                {\tt \{T t = e1; t op2= e2; x op1= t;\}}    \`(7)\\\\

{\tt \{T t1 = e1; s1;\}\{T t2 = e2; s2;\}} \,$\equiv$\, \\
                {\tt \{T t = e1; s1; t = e2; s2;\}}.        \`(8)
\end{tabbing}
Here {\tt x}, {\tt t}, {\tt t1}, {\tt t2} are variables of type {\tt
T}; {\tt e1}, {\tt e2} are expressions; and self-mutating
operators {\tt op=}, {\tt op1=}, {\tt op2=} correspond to operators
{\tt op}, {\tt op1}, {\tt op2}, respectively.  Recall that Sophus does
not allow aliasing. Rule (7) introduces a temporary variable
{\tt t} in a local environment and rule (8) reduces the number of
temporary variables by merging two local environments declaring a
temporary into a single one.

\subsection{Benchmarks} \label{sec:BENCHMARKS}

\subsubsection{Two Kernels}

\begin{figure}
\rule{7.5cm}{.5mm}
\begin{verbatim}
template<class T> void F (T & x)
{ x = x*x + x*2.0; }

template<class T> void P (T & x)
{ T temp1 = x;
  temp1 *= 2.0 ;
  x *= x;
  x += temp1;
}
\end{verbatim}
\rule{7.5cm}{.5mm}
\caption{Kernels \texttt{F} and \texttt{P}.} \label{fig:QUAD}
\end{figure}

Consider C++ procedures \texttt{F} and \texttt{P} shown in
Figure~\ref{fig:QUAD}.  \texttt{F} computes $x^2+2x$ using algebraic
notation while \texttt{P} computes the same expression in
self-mutating form using a single temporary variable {\tt temp1}.
Both were run with meshes of different sizes.  The corresponding
timing results are shown in
Figures~\ref{fig:benchmark-sun},~\ref{fig:benchmark-cray},
and~\ref{fig:benchmark-sunsmall}.

\begin{figure*}[tp]
\begin{center}\small
\begin{tabular}{|r@{}l|c||r|r|r||r|r|r||r|r|} \hline
&&&\multicolumn{8}{c|}{SUN Ultra-2}\\
\cline{4-11}
        \multicolumn{2}{|c|}{Number of} &  Type & 
        \multicolumn{3}{c||}{No options} & \multicolumn{3}{c||}{Option {\tt -fast}} & 
        \multicolumn{2}{c|}{Optim. speedup}
        \\
\multicolumn{2}{|c|}{elements}  &       
                            & NC & NS & NS/NC 
                                             & OC & OS & OS/OC &NC/OC&NS/OS \\\hline\hline
               &            & {\tt F} & 6.4s & 28.4s & 4.4 & 2.8s & 4.7s & 1.7 & 2.3 & 6.0 \\
  $8^{3}= 2$&kB       & {\tt P} & 7.8s & 12.2s & 1.6 & 2.8s & 2.0s & 0.7 & 1.8 & 6.1 \\
               &            &{\tt F}/{\tt P}& 0.8 &  2.3 &     & 1.0 & 2.4 &     &     & \\\hline\hline
               &            & {\tt F} & 6.8s & 31.7s & 4.7 & 3.2s & 8.3s & 2.6 & 2.1 & 3.8 \\
 $64^{3}= 1$&MB   & {\tt P} & 8.2s & 13.4s & 1.6 & 3.2s & 3.4s & 1.1 & 2.6 & 3.9 \\
               &            &{\tt F}/{\tt P}& 0.8 &  2.4 &     & 1.0 & 2.4 &     &     & \\\hline\hline
               &            & {\tt F} & 7.1s & 238.5s & 33.6 & 3.5s &199.3s& 56.9 & 2.0 & 1.2 \\
$256^{3}= 6$&$7$MB & {\tt P} & 8.5s & 15.6s & 1.8 & 3.5s & 18.5s & 5.3 & 2.4 & 0.8 \\
               &            &{\tt F}/{\tt P}& 0.8 & 15.3 &     & 1.0 & 10.8& & & \\\hline
\end{tabular}
\end{center}
\caption{Speed of conventional vs. Sophus style on SUN sparc 
Ultra-2 workstation. More specifically, a \emph{SunOS 5.6 Generic\_105181-06
sun4u sparc SUNW, Ultra-2} hardware platform with 512MB internal memory
and the SunSoft C++ compiler \emph{CC: WorkShop Compilers 4.2 30 Oct
1996 C++ 4.2} were used.
}
\label{fig:benchmark-sun}
\end{figure*}
\begin{figure*}[tp]
\begin{center}\small
\begin{tabular}{|r@{}l|c||r|r|r||r|r|r||r|r|} \hline
&&&\multicolumn{8}{c|}{Silicon Graphics/Cray Origin 2000}\\
\cline{4-11}
        \multicolumn{2}{|c|}{Number of} &  Type & 
        \multicolumn{3}{c||}{No options} & \multicolumn{3}{c||}{Option {\tt -Ofast}} & 
        \multicolumn{2}{c|}{Optim. speedup}
        \\
\multicolumn{2}{|c|}{elements}  &       
                            & NC & NS & NS/NC 
                                             & OC & OS & OS/OC &NC/OC&NS/OS \\\hline\hline
               &            & {\tt F} & 3.3s & 10.5s & 3.2 & 1.0s & 2.3s & 2.3 & 3.3 & 4.6 \\
  $8^{3}= 2$&kB       & {\tt P} & 4.4s &  6.3s & 1.4 & 1.0s & 1.3s & 1.3 & 4.4 & 4.8 \\
               &            &{\tt F}/{\tt P}& 0.8 &  1.6 &     & 1.0 & 1.8 &     &     & \\\hline\hline
               &            & {\tt F} & 3.4s & 12.3s & 3.6 & 1.3s & 4.1s & 3.2 & 2.6 & 3.0 \\
 $64^{3}= 1$&MB   & {\tt P} & 4.5s &  6.9s & 1.5 & 1.2s & 2.3s & 1.9 & 3.8 & 3.0 \\
               &            &{\tt F}/{\tt P}& 0.8 &  1.8 &     & 1.1 & 1.8 &     &     & \\\hline\hline
               &            & {\tt F} & 4.0s & 25.0s & 6.3 & 1.8s &15.6s & 8.7 & 2.2 & 1.6 \\
$256^{3}= 6$&$7$MB & {\tt P} & 5.2s & 10.3s & 2.0 & 1.7s & 7.0s & 4.1 & 3.1 & 1.5 \\
               &            &{\tt F}/{\tt P}& 0.8 &  2.4 &     & 1.1 & 2.2 &     &     & \\\hline
\end{tabular}
\end{center}
\caption{Speed of conventional vs. Sophus style on Silicon Graphics/Cray Origin 2000.
More specifically, the Origin 2000 had hardware version
\emph{IRIX64 ask 6.5SE 03250013 IP27} with a total of 24GB memory
distributed among 128 processors. The C++ compiler used was
\emph{MIPSpro Compilers: Version 7.2.1.1m}.
}
\label{fig:benchmark-cray}
\end{figure*}
\begin{figure*}[tp]
\begin{center}\small
\begin{tabular}{|r@{}l|c||r|r|r||r|r|r||r|r|} \hline
&&&\multicolumn{8}{c|}{SUN Ultra-2}\\
\cline{4-11}
        \multicolumn{2}{|c|}{Number of} &  Type & 
        \multicolumn{3}{c||}{No options} & \multicolumn{3}{c||}{Option {\tt -fast}} & 
        \multicolumn{2}{c|}{Optim. speedup}
        \\
\multicolumn{2}{|c|}{elements}  &       
                            & NC & NS & NS/NC 
                                             & OC & OS & OS/OC &NC/OC&NS/OS \\\hline\hline
               &            & {\tt F} & 6.4s & 28.4s & 4.4 & 2.9s & 4.7s & 1.6 & 2.2 & 6.0 \\
 $8^{3}= 2$&kB        & {\tt P} & 7.8s & 12.2s & 1.6 & 2.8s & 2.0s & 0.7 & 2.8 & 6.1 \\
               &            &{\tt F}/{\tt P}& 0.8 &  2.3 &     & 1.0 & 2.4 &     &     & \\\hline\hline
               &            & {\tt F} & 6.5s & 28.9s & 4.4 & 3.2s & 5.2s & 1.6 & 2.0 & 5.6 \\
$16^{3}= 1$&$6$kB  & {\tt P} & 7.9s & 12.6s & 1.6 & 3.2s & 2.5s & 0.8 & 2.5 & 5.0 \\
               &            &{\tt F}/{\tt P}& 0.8 &  2.3 &     & 1.0 & 2.1 &     &     & \\\hline\hline
               &            & {\tt F} & 6.8s & 29.5s & 4.3 & 3.2s & 6.2s & 1.9 & 2.1 & 4.8 \\
$32^{3}= 1$&$28$kB   & {\tt P} & 8.1s & 12.8s & 1.6 & 3.1s & 2.7s & 0.9 & 2.6 & 4.7 \\
               &            &{\tt F}/{\tt P}& 0.8 &  2.3 &     & 1.0 & 2.3 &     &     & \\\hline\hline
               &            & {\tt F} & 6.8s & 31.7s & 4.7 & 3.2s & 8.3s & 2.6 & 2.1 & 3.8 \\
$64^{3}= 1$&MB    & {\tt P} & 8.2s & 13.4s & 1.6 & 3.2s & 3.4s & 1.1 & 2.6 & 3.9 \\
               &            &{\tt F}/{\tt P}& 0.8 &  2.4 &     & 1.0 & 2.4 &     &     & \\\hline
\end{tabular}
\end{center}
\caption{Speed of conventional vs. Sophus style on SUN sparc 
Ultra-2 workstation for small meshes.}
\label{fig:benchmark-sunsmall}
\end{figure*}

The mesh size is given in the leftmost column. Mesh elements are
single precision reals of 4B each. The second column indicates the
benchmark procedure ({\tt F} or {\tt P}) or the ratio of the
corresponding timings ({\tt F}/{\tt P}).  The columns NC, NS, OC, and
OS give the time in seconds of several iterations over each mesh so
that a total of 16\,777\,216 elements were updated in each case. This
corresponds to $32\,768$ iterations for mesh size $8^3$, $64$
iterations for mesh size $64^3$, $1$ iteration for mesh size $256^3$,
and so forth.  In columns \_C (conventional style) the procedure calls
are performed for each element of the mesh, 
while in columns \_S (Sophus style) they are performed as operations
on the entire mesh variables.

Columns N\_ give the time for unoptimized code (no compiler options),
while columns O\_ give the time for code optimized for speed (compiler
option {\tt -fast} for the SUN CC compiler and {\tt -Ofast} for the
Silicon Graphics/Cray CC compiler).  The timings represent the median of 5  test
runs. These turned out to be relatively stable measurements, except in
columns NS and OS, rows $256^3$ {\tt F} and {\tt P} of
Figure~\ref{fig:benchmark-sun}, where the time for an experiment could
vary by more than 100\%. This is probably due to paging activity on
disk dominating the actual CPU time. Also note that the
transformations done by the optimizer are counterproductive in the
{\tt P} case, yielding an NS/OS ratio of 0.8.

When run on the SUN the tests where the only really active processes,
while the Cray was run in its normal multi-user mode but at a
relatively quiet time of the day (Figure~\ref{fig:cray-use}).  As can
be seen the load was moderate (around 58) and although fully utilized,
resources where not overloaded.

\begin{figure}
\rule{7.5cm}{.5mm}
\footnotesize
\begin{verbatim}
IRIX64 ask 6.5SE IP27          
load averages: 58.37 57.74 58.30     06:46:21
385 processes: 323 sleeping, 3 stopped,
               1 ready, 58 running
128 CPUs:  0.0% idle, 0.0% usr,  0.0% ker,
           0.0% wait, 0.0% xbrk, 0.0% intr
Memory: 24G max,  23G avail, 709M free,
        25G swap, 17G free swap
\end{verbatim}
\normalsize
\rule{7.5cm}{.5mm}
\caption{Random load information for test run on Silicon Graphics/Cray Origin 2000.}
\label{fig:cray-use}
\end{figure}

In the current context, only columns NS and OS are relevant, the other
ones are explained in Section~\ref{sec:INEFF}.  As expected, the
self-mutative form {\tt P} is a better performer than the algebraic
form {\tt F} when the Sophus style is used.  Disregarding the cases
with disk paging mentioned above, we see that the self-mutating mesh
operations are 1.8--2.4 times faster than their algebraic
counterparts, i.e., the CodeBoost transformation roughly doubles the
speed of these benchmarks.

\subsubsection{Full Application: SeisMod} \label{sec:SEISMOD}

We also obtained preliminary results on the Silicon Graphics/Cray Origin 2000 for a
full application, the seismic simulation code SeisMod, which is
written in C++ using the Sophus style. It is a collection of
applications using the finite difference method for seismic
simulation. Specific versions of SeisMod have been tailored to handle
simulations with simple or very complex geophysical
properties.\footnote{SeisMod is used and licensed by the geophysical
modelling company UniGEO A.S. (Bergen, Norway).} We compared a version of
SeisMod implemented using SCC generated wrapper functions and a
self-mutating version produced by the CodeBoost source-to-source
optimizer:
\begin{itemize}

\item   The algebraic expression style version turned out to give
  a 10--30\% reduction in source code size and greatly enhanced
  readability for complicated parts of the code.  This implies a
  significant programmer productivity gain as well as a significant
  reduction in maintenance cost as measured by the Cocomo technique
  \cite{Boehm81}, for instance

\item   A 30\% speed increase was obtained after 10 selected procedures out
   of 150 procedures with speedup potential had been brought in
   self-mutating form.  This speedup turned out to be independent of
   C++ compiler optimization flag settings.

\end{itemize}

This shows that although a more user-friendly style may give a
performance penalty compared to a conventional style, it is possible
to regain much of the efficiency loss by using appropriate
optimization tools. Also, a more abstract style may yield more
cost-effective software, even without these optimizations, if the
resulting development and maintenance productivity improvement is
taken into account.


\section{Implementation of CodeBoost} \label{sec:CODEBOOST}

CodeBoost is a dedicated C++ source-to-source transformation tool for
Sophus style programs.  It has been implemented using the ASF+SDF
language prototyping system \cite{DHK96}.  ASF+SDF allows the required
transformations to be entered directly as conditional rewrite rules
whose right- and left-hand sides consist of language (in our case C++)
patterns with variables and auxiliary transformation functions.  The
required language specific parsing, rewriting, and prettyprinting
machinery is generated automatically by the system from the high-level
specification.  Program transformation tools for Prolog and the
functional language Clean implemented in ASF+SDF are described in
\cite{Brunekreef96,vandenBrandEtAl95}.

An alternative implementation tool would have been the TAMPR program
transformation system \cite{Boyle89}, which has been used successfully
in various HPC applications.  We preferred ASF+SDF mainly because of
its strong syntactic capabilities enabling us to generate a C++
environment fairly quickly given the complexity of the
language.

Another alternative would have been the use of template metaprogramming
and/or expression templates \cite{Veldhuizen95,VeldhuizenJernigan97}.
This approach is highly C++ specific, however, and cannot be adapted
to Fortran 90/95.

Basically, the ASF+SDF implementation of CodeBoost involves the
following two steps:
\begin{enumerate}

\item Specify the C++ syntax in SDF, the syntax definition
      formalism of the system.\\

\item Specify the required transformation rules as conditional rewrite
      rules using the C++ syntax, variables, and auxiliary
      transformation functions.

\end{enumerate}

\begin{figure*}
\rule{15.8cm}{.5mm}
\setlength{\unitlength}{3750sp}%
\begingroup\makeatletter\ifx\SetFigFont\undefined%
\gdef\SetFigFont#1#2#3#4#5{%
  \reset@font\fontsize{#1}{#2pt}%
  \fontfamily{#3}\fontseries{#4}\fontshape{#5}%
  \selectfont}%
\fi\endgroup%
\begin{picture}(5850,4344)(376,-4390)
\put(1275,-2000){\makebox(0,0)[lb]{\smash{\SetFigFont{11}{13.2}{\rmdefault}{\mddefault}{\updefault}Macro expansion}}}
\put(1150,-2225){\makebox(0,0)[lb]{\smash{\SetFigFont{11}{13.2}{\rmdefault}{\mddefault}{\updefault}(C++ preprocessor)}}}
\put(6450,-4111){\makebox(0,0)[lb]{\smash{\SetFigFont{11}{13.2}{\rmdefault}{\mddefault}{\updefault}Optimized C++}}}
\put(6750,-4336){\makebox(0,0)[lb]{\smash{\SetFigFont{11}{13.2}{\rmdefault}{\mddefault}{\updefault}program}}}
%
\thicklines
\put(3076,-3211){\oval(600,1500)[tl]}
\put(3076,-3211){\oval(600,1500)[bl]}
\put(5401,-3211){\oval(600,1500)[br]}
\put(5401,-3211){\oval(600,1500)[tr]}
\put(1100,-886){\vector( 0,-1){3000}}
\put(1951,-4080){\vector( 1, 0){4400}}
\put(1600,-660){\line( 1,-0){2550}}
\put(4150,-810){\oval(300,300)[tr]}
\put(4300,-810){\vector( 0,-1){1550}}
\put(1576,-61){\makebox(0,0)[lb]{\smash{\SetFigFont{11}{13.2}{\rmdefault}{\mddefault}{\updefault}    }}}
\put(576,-4111){\makebox(0,0)[lb]{\smash{\SetFigFont{11}{13.2}{\rmdefault}{\mddefault}{\updefault}C++ program}}}
\put(376,-4336){\makebox(0,0)[lb]{\smash{\SetFigFont{11}{13.2}{\rmdefault}{\mddefault}{\updefault}(macros expanded)}}}
\put(526,-575){\makebox(0,0)[lb]{\smash{\SetFigFont{11}{13.2}{\rmdefault}{\mddefault}{\updefault}Sophus C++}}}
\put(676,-800){\makebox(0,0)[lb]{\smash{\SetFigFont{11}{13.2}{\rmdefault}{\mddefault}{\updefault}program}}}
\put(4350,-1560){\makebox(0,0)[lb]{\smash{\SetFigFont{11}{13.2}{\rmdefault}{\mddefault}{\updefault}Perl script}}}
\put(3625,-2600){\makebox(0,0)[lb]{\smash{\SetFigFont{11}{13.2}{\rmdefault}{\mddefault}{\updefault}Declarations in}}}
\put(3350,-2825){\makebox(0,0)[lb]{\smash{\SetFigFont{11}{13.2}{\rmdefault}{\mddefault}{\updefault}Sophus C++ program}}}
\put(3075,-3050){\makebox(0,0)[lb]{\smash{\SetFigFont{11}{13.2}{\rmdefault}{\mddefault}{\updefault}(as ASF+SDF specification)}}}
\put(4275,-3350){\makebox(0,0)[b]{\smash{\SetFigFont{11}{13.2}{\rmdefault}{\bfdefault}{\updefault}specializes}}}
\put(3275,-3650){\makebox(0,0)[lb]{\smash{\SetFigFont{11}{13.2}{\rmdefault}{\mddefault}{\updefault}CodeBooost ASF+SDF}}}
\put(3775,-3875){\makebox(0,0)[lb]{\smash{\SetFigFont{11}{13.2}{\rmdefault}{\mddefault}{\updefault}specification}}}
\end{picture}
\rule{15.8cm}{.5mm}
\caption{Two-phase specification of CodeBoost.} \label{fig:TWOPHASE}
\end{figure*}
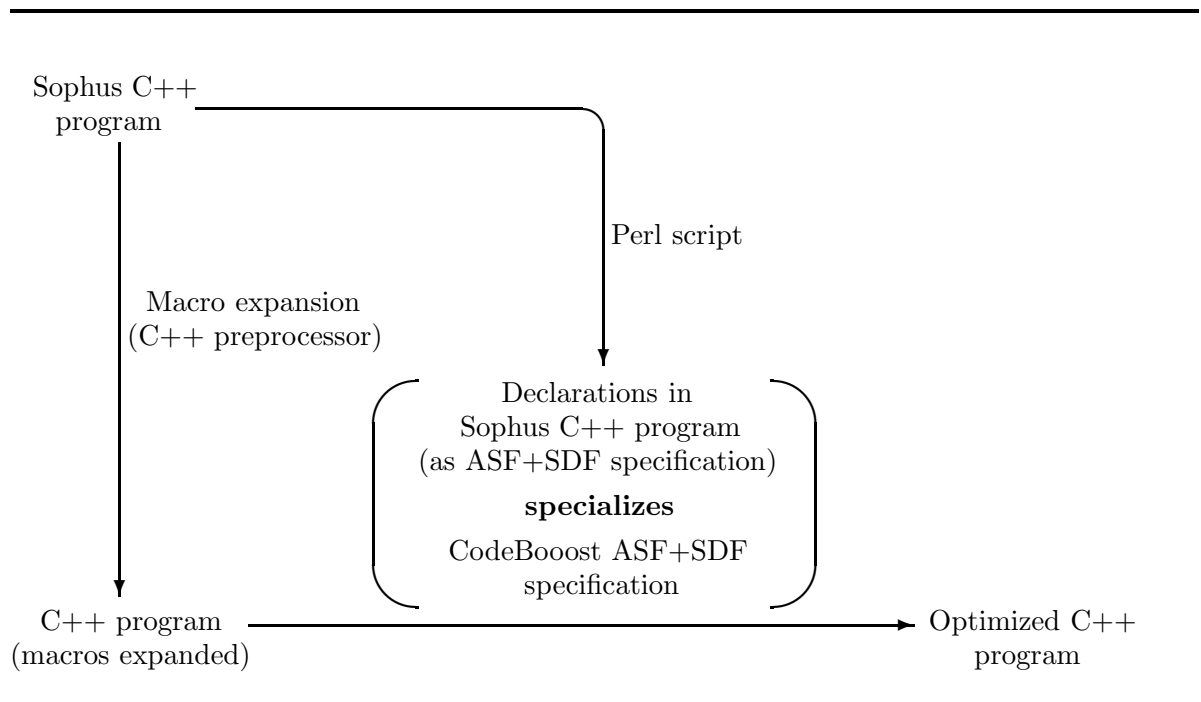

As far as the first step is concerned, specification of the large C++
syntax in SDF would involve a considerable effort, but fortunately a
BNF-like version is available from the ANSI C++ standards committee.
We obtained a machine-readable preliminary version
\cite{ATT96} and translated it largely automatically into SDF
format. The ASF+SDF language prototyping system then generated a C++
parser from it. The fact that the system accepts general context-free
syntax rather then only LALR or other restricted forms of syntax also
saved a lot of work in this phase even though the size of the C++
syntax taxed its capabilities.

With the C++ parser in place, the required program transformation
rules were entered as conditional rewrite rules.  In general, a
program transformer has to traverse the syntax tree of the program to
collect the context-specific information used by the actual
transformations.  In our case, the transformer needs to collect the
declaration information indicating which of the operations have a
self-mutating implementation. Also, in Sophus the self-mutating
implementation of an operator (if any) need not update {\it this} but
can indicate which of the arguments is updated using the {\tt upd}
flag. The transformer therefore needs to collect not only which of the
operations have a self-mutating implementation but also which argument
is being mutated in each case.  As a consequence, CodeBoost has to
traverse the program twice: once to collect the declaration
information and a second time to perform the actual
transformations. Two other issues have to be taken into account:
\begin{itemize}

\item C++ programs cannot be parsed before their macros are
 expanded. Some Sophus-specific language elements are implemented as
 macros, but are more easily recognized before expansion than after.
 An example is the {\tt upd} flag indicating which argument of an
 operator or function is the one to be updated.

\item Compared to the total number of constructs in C++,
 there are relatively few constructs of interest.  Since all
 constructs have to be traversed, this leads to a plethora of trivial
 tree traversal rules. As a result, the specification gets cluttered
 up by traversal rules, making it a lot of work to write as well as
 hard to understand.  One would like to minimize or automatically
 generate the part of the specification concerned with straightforward
 program traversal.

\end{itemize}

Our approach to the above problems is to give the specification a
two-phase structure as shown in Figure~\ref{fig:TWOPHASE}.  Under the
reasonable assumption that the declarations are not spoiled by macros,
the first phase processes the declarations of interest prior to macro
expansion using a stripped version of the C++ grammar that captures
the declaration syntax only.  We actually used a Perl script for this,
but it could have been done in ASF+SDF as well.  It yields an ASF+SDF
specification that is added to the specification of the second phase.
The effect of this is that the second phase is specialized for the
program at hand in the sense that the transformation rules in the
second phase can assume the availability of the declaration
information and thus can be specified in a more algebraic, i.e.,
context independent manner.  As a consequence, they are easy to read,
consisting simply of the rules for the constructs that may need
transformation and using the ASF+SDF system's built-in innermost tree
traversal mechanism.  In this way, we circumvented the last-mentioned
problem.

As CodeBoost is developed further, it will have to duplicate more and
more functions already performed by any C++ preprocessor/compiler.
Not only will it have to do parsing (which it is already doing now),
but also template expansion, overloading resolution, and dependence
analysis.  It would be helpful if CodeBoost could tap into an existing
compiler at appropriate points rather than redo everything itself. One
of the candidates we are considering is the IBM Montana C++
compiler/programming environment
\cite{SorokerEtAL97}, which provides an open architecture with APIs
giving access to various compiler intermediate representations with
pointers back to the source text.

\section{Software Structure vs. Efficiency} \label{sec:SOFTWARESTRUCT}

As noted in Section~\ref{sec:INTRO}, programs in a domain-specific
programming style like Sophus may need additional optimization in view
of their increased use of expensive constructs. On the other hand, the
restrictions imposed by the style may lead to new high-level
optimization opportunities that can be exploited by a CodeBoost-like
optimization tool.  We give some further examples of both phenomena.

\subsection{Inefficiencies Caused by the Use of an Abstract Style} \label{sec:INEFF}

We consider an example. As explained in Section~\ref{sec:SOPHUS},
scalar field operations like $+$ and $*$ are implemented on top of
mesh operations $+$ and $*$. The latter will typically be implemented
as iterations over all array elements, performing the appropriate operations
pairwise on the elements. For scalar fields, expressions like

\noindent $X_{1} = A_{1,1}*V_{1}+A_{1,2}*V_{2}$,\\
$X_{2} = A_{2,1}*V_{1}+A_{2,2}*V_{2}$

\noindent will force 8 traversals over the mesh data structure. If the
underlying meshes are large, this may cause many cache misses for each
traversal. Now each of the scalar fields $A_{i,j}$, $V_{j}$, and
$X_{j}$ are actually implemented using a mesh, i.e., an array of $n$
elements, and are represented in the machine by {\tt A[i,j,k]}, {\tt
V[j,k]} and {\tt X[j,k]} for $\mathtt{k}=1,\ldots,\mathtt{K}$, where
{\tt K} is the number of mesh points of the discretisation. In a
conventional implementation this would be explicit in the code more or
less as follows:
\begin{verbatim}
for k := 1,K
    for j := 1,2
        X[j,k] := 0
        for i := 1,2
            X[j,k] += A[i,j,k]*V[j,k]
        endfor
    endfor
endfor
\end{verbatim}
It would be easy for an optimizer to partition the loops in such a way
that the number of cache misses is reduced by a factor of 8.

In the abstract case aggressive in-lining is necessary to expose the
actual loop nesting to the optimizer.  Even though most existing C++
compilers do in-lining of abstractions, this would be non-trivial
since many abstraction layers are involved from the programmer's
notation on top of the library of abstractions down to the actual
traversals being performed.

Consider once again the timing results shown in
Figure~\ref{fig:benchmark-sun}, Figure~\ref{fig:benchmark-cray}, and
Figure~\ref{fig:benchmark-sunsmall}. As was explained in
Section~\ref{sec:BENCHMARKS}, the procedure calls in columns \_C
(conventional style) are performed for each element of the mesh, while
they are performed as operations on the entire mesh variables in
columns \_S (Sophus style).  Columns OS/OC for row {\tt P} give the
relevant figures for the performance loss of optimized Sophus style
code relative to optimized conventional style code as a result of
Sophus operating at the mesh level rather than at the element
level. The benchmarks show a penalty of 1.1--5.3, except for data
structures of less than 128kB on the SUN, where a speedup of up to 1.4
(penalty of 0.7) can be seen in Figure~\ref{fig:benchmark-sunsmall}.
As is to be expected, for large data structures the procedure calls in
column OC are more efficient than those in column OS, as the optimizer
is geared towards improving the conventional kind of code consisting
of large loops with procedure calls on small components of data
structures. Also, cache and memory misses become very costly when
large data structures have to be traversed many times.

The figures for {\tt P} in column OS of
Figure~\ref{fig:benchmark-sunsmall} are somewhat unexpected.  In these
cases OS is the fastest alternative up to a mesh size somewhere
between $32^3$ and $64^3$. This may be due to the smaller number of
procedure calls in the OS case than in the OC case. In the latter case
{\tt F} and {\tt P} are called once per element, i.e., 16\,777\,216
times, while in the OS case they are called only once and the
self-mutating operations are called only 4 times.

Another interesting phenomenon can be seen in column NC of
Figure~\ref{fig:benchmark-sun} and
Figure~\ref{fig:benchmark-cray}. Here the self-mutating version takes
longer than the algebraic version, probably because the compiler
automatically puts small temporaries in registers for algebraic
expressions, but cannot do so for self-mutating forms. The OC column
shows that the optimizer eliminates the difference.



\subsection{New Opportunities for Optimization} \label{sec:OPPORT}

The same abstractions that were a source of worry in the previous
section at the same time provide the specificity and typing making the
use of high-level optimizations possible.  Before they are removed by
inlining, the information the abstractions provide can be used to
select and apply appropriate datatype specific optimization rules.
Sophus allows application of such rules at very high levels of
abstraction.  Apart from the expression transformation rules (1)--(8)
(Section~\ref{sec:OPNOTATION}), which are applicable to a wide range
of operators and functions, further examples at various levels of
abstraction are:
\begin{itemize}
\item The laws of tensor algebra.  In Sophus the tensors contain the
      continuous scalar fields as elements (Section~\ref{sec:SOPHUS}),
      thus making the abstract tensor operations explicit in appropriate
      modules.

\item Specialization of general tensor code for specific coordinate
      systems. A Cartesian coordinate system gives excellent
      simplification and axiosymmetric ones also give good
      simplification compared to general curvilinear code.

\item Optimization of operations on matrices with many symmetries.
      Such symmetries offer opportunities for optimization in many
      cases, including the seismic modelling application mentioned in
      Section~\ref{sec:SEISMOD}.
\end{itemize}


\section{Conclusions and Future Work} \label{sec:FINAL}

\begin{itemize}

\item The Sophus class library in conjunction with 
      the CodeBoost expression transformation tool shows the
      feasibility of a style of programming PDE solvers that attempts
      to stay close to the abstract mathematical theory in terms of
      both the datatypes and the algebraic style of expressions used.

\item Our preliminary findings for a full application, the Sophus style
      seismic simulation code SeisMod, indicate significant programmer
      productivity gains as a result of adopting the Sophus style.

\item There are numerous further opportunities for optimization by
      CodeBoost in addition to replacement of appropriate operators
      and functions by their self-mutating versions. Sophus allows
      datatype specific rules to be applied at very high levels of
      abstraction.

\end{itemize}

\section*{Acknowledgments}
Hans Munthe-Kaas, Andr\'e Friis, Kristin Fr{\o}ysa, Steinar S{\o}reide, and
Helge Gunnarsli have contributed to Sophus in various ways.   

\bibliographystyle{abbrv}

\end{document}